\documentclass[12pt]{iopart}
\usepackage{graphicx}	

\usepackage{iopams}
\newcommand{\bm}[1]{\mbox{\boldmath$#1\!$}}
\newcommand{\bn}[1]{\mbox{\boldmath$#1$}}

\newcommand{\beq}{\begin{equation}}
\newcommand{\eeq}{\end{equation}}
\newcommand{\bea}{\begin{eqnarray}}
\newcommand{\eea}{\end{eqnarray}}

\begin{document}
\centerline{(Authors' post-acceptance version of the paper published in J. Opt. {\bf 21} 104002, 23 Sept 2019)}

\title[J. Opt {\bf 21} (2019) 104002 https//doi.org/10.1088/2040-8996/ab411f]{Interference of axially-shifted Laguerre-Gaussian beams and their interaction with atoms}
\author{K. Koksal$^1,3$, V. E. Lembessis$^2$, J. Yuan$^3$ and M. Babiker$^3$\footnote{Author to whom any correspondence should be addressed}}
\address{$^1$ Department of Physics, Bitlis Eren University, 130000 Bitlis, Turkey}
\address{$^2$ Quantum Technology Group, Department of Physics and Astronomy, College of Science, King Saud University, P O Box 2455, Riyadh 11451, Saudi Arabia}
\address{$^3$ Department of Physics, University of York, Heslington, York YO10 5DD, England,  UK}
\ead{meb6@york.ac.uk}
\begin{abstract}
Counter-propagating co-axial Laguerre-Gaussian (LG) beams are considered, not in the familiar scenario where the focal planes coincide at $z=0$, but when they are separated by a finite axial distance $d$. The simplest case is where both beams are doughnut beams which have the same linear polarisation.  The total fields of this system are shown to display novel amplitude and phase distributions and are shown to give rise to a ring or a finite ring lattice composed of double rings and single central ring.  When the beams have slightly different frequencies the ring lattice pattern becomes a finite set of rotating Ferris wheels and the whole pattern also moves axially between the focal planes.  We show that the field of such an axially shifted pair of counter-propagating LG beams generate trapping potentials due to the dipole force which can trap two-level atoms in the components of the ring lattice.  We also highlight a unique feature of this system which involves the creation of a new longitudinal optical atom trapping potential due to the scattering force which arises solely when $d \ne 0$. The results are illustrated using realistic parameters which also confirm the importance of the Gouy and curvature effects in determining the ring separation both radially and axially and gives rise to the possibility of atom tunnelling between components of the double rings.
\end{abstract}


\centerline{({\bf PACS Numbers: 42.50Tx; 42.50.Wk; 78.68.+m; 37.10.De; 37.10.Vz})}

\section{Introduction}

Optical manipulation as an area of optical physics stems from the mechanical effects which laser light imposes on matter, both in the form of atoms and molecules or matter at the nanoscale.  This area continues to be a subject of much interest both from a fundamental point of view and for useful applications. Research began with the pioneering work by Ashkin et al \cite{Ashkin1986} who showed that light pressure applied to atoms, leads to a variety of techniques resulting, most notably,  in their heating, cooling,  trapping and levitation (see also \cite{Chu1986}. The prototypical example of optical manipulation is the `optical tweezers’ \cite{Ashkin1970} and their are recent accounts involving interferometric multi-beam and surface optical tweezers \cite{Mohammadnezhad2017a, Mohammadnezhad2017b}. However, the most celebrated manifestation of radiation pressure is in the realisation of Bose-Einstein condensation in dilute atomic systems \cite{Anderson1995}.

Recently, optical manipulation received an impetus when combined with structured laser light in which optical fields can have unlimited possibilities of form in terms of spatial and phase variations \cite{Andrews2008a}. Laser beams in arbitrary geometrical arrangements can be made to interfere in a specified fashion and so generate optical potential landscapes and associated forces and torques leading to atom manipulation.  Ordered arrangements of interfering laser beams form the so-called optical lattices as an example of structured light \cite{Jessen1992, Verkerk1992}.

More recently, a new ingredient entered the arena of optical manipulation with the discovery of vortex laser light, which is light endowed with the property of orbital angular momentum (OAM)\cite{Allen1992}. Such light beams are exemplified by the Laguerre-Gaussian (LG) beams distinguished by the two indices $l,p$ and carry OAM of magnitude $l\hbar$ per photon.  The OAM property of this type of laser light is in addition to the wave polarisation, or optical spin angular momentum (SAM) property, which is intrinsic to all laser light (see \cite{Torres2011} to \cite{Barnett2017}). In this context both atomic and nanoparticle manipulation have revealed additional effects, with the OAM features manifesting themselves as rotational motion, and new atom trapping and cooling techniques have emerged (see \cite{Babiker2018,He1995, Friese1998, Clifford1998, Dholakia2002, Grier2003, Ladavac2004, Galajda2001, Vickers2008, Franke-Arnold2007, Anderson2006, Lembessis2011, Lembessis2016, Hayrapetyan2013, Rsheed2016}). The optical spanner is one of the first applications which has been realised as the rotational form of the optical tweezers (see \cite{He1995, Friese1998, Clifford1998, Dholakia2002, Grier2003, Ladavac2004, Galajda2001}).

The simplest scenario consists a single focused beam where typically the scattering force attracts the atoms towards the focal plane where the beam intensity is maximum.  However an axial force in the direction of beam propagation leads to axial drifting which has been counteracted with the use of  counter-propagating beams \cite{Vickers2008}. Counter-propagating LG beams with coinciding focal planes have been considered as a means of localising cold atoms in the dark regions of the total fields \cite{Arnold2012}. It has also been shown that when two vortex beams meet at their common focal plane, the interference results in a  petal-like intensity pattern and such a petal pattern rotates when the two beams are slightly different in frequency, resulting in the Ferris-wheel phenomenon which was discussed first by Franke-Arnold et al \cite{Franke-Arnold2007} and later by Vickers \cite{Vickers2008}.   The case of two counter-propagating Laguerre-Gaussian doughnut beams LG$_{l,0}$ and LG$_{-l,0}$ with orthogonal linear polarizations ${\bn {\hat {\epsilon}}}_x$ and ${\bn {\hat {\epsilon}}}_y$ leads to azimuthal polarization gradients and an azimuthal Sisyphus effect that can be utilized in the creation and control of a persistent current of superfluid atoms circulating in a toroidal trap (see \cite{Anderson2006, Lembessis2011, Lembessis2016}).  In addition to the case of LG beams, the coupling of atoms to Bessel beams has been considered \cite{Hayrapetyan2013}.

In this paper we consider a scenario of interference which, as far as the authors know,  has not been explored before in the context of optical vortices.  This simply involves the introduction of a finite spatial separation $d$ between the two focal planes of the counter-propagating LG beams, giving  rise to complex intensity and phase changes. The simplest case is where both beams are doughnut beams and have the same linear polarisation ${\bn {\hat {\epsilon}}}$. The prominent feature of the resulting intensity distribution is the overlap region of the two beams, the size of which can be controlled by changing the separation $d$ and the curvature of the beams which alter the useful length of the interference pattern.  Besides the intensity, which is shown to be in the form of a finite ring lattice, new features arise due to phase variations in the phase function and its gradient.  The total fields of this system are shown to display novel amplitude and phase distributions and are shown to give rise to a finite ring lattice composed of double rings and single central ring.  When the beams have slightly different frequencies the interference pattern becomes a set of rotating Ferris wheels and the whole pattern also moves axially between the focal planes.  

This papers is organised as follows.  We firstly display the standard formalism of LG beams which is essential for the theoretical development that follows.  We begin with the evaluation of the case in which the two LG beams both have low-intensity.  We show that the scattering forces exhibit an additional axial component between the focal planes which is due to the axial shift and points symmetrically towards the centre of the beams.  This is a new form of optical trapping of atoms based on scattering forces and arising solely due to the axial shift.  Next, we deal with the general case of higher light intensities and large winding number $l$.  Here, the full formalism of the interference has to be deployed, along with numerical analysis.  A prominent feature of the general theory for $d \ne 0$ is the inter-mixing of the phase and amplitudes due to the shifted beams in the total fields. The full theory enables the realisations of ring Ferris wheels and axial pattern motion to be manifested between the focal planes when the shifted beams have slightly different frequencies.  

The main results of this papers can be summarised as follows. (1) We identify and quantify a new trapping mechanism for two-level atoms that arises due to the scattering forces which vanishes in the more common cases with $d=0$. We feel that this result is significant, not only because of the novelty of the finding, but also because it entails an all-optical trapping mechanism that would be an alternative to other more involved trapping mechanisms such as magneto-optical trapping. (2) The one-dimensional case considered here highlights the role of the Gouy phase and curvature in that they control the spacing and number of rings in the interference pattern.  We show that the phase difference controls various features including the number and spacing (both radially and axially) of the rings. (3) We also point out and quantify effects that would be of interest to experimentalists, including atom trapping by both the scattering force and the dipole force and we examine regimes for atom tunnelling between double rings in the finite lattice.

\section{LG optical modes}
The electric field vector distribution of a single LG mode, characterised by the integers $l$ and $p$, of frequency $\omega$ and axial wavevector $k$ travelling along the positive z-axis, can be written in cylindrical coordinates ${\bf r}=(r,\phi,z)$ as follows
\beq
{\bf E}_{klp}(\rho,\phi,z)=U_{klp}e^{i\Theta_{klp}(\rho,\phi,z)}e^{i\omega t}{\bn {\hat {\epsilon}}},
\eeq
where ${\bn {\hat {\epsilon}}}$ is the wave polarisation vector, $U_{klp}(r,\phi,z)$ is the amplitude function
and $\Theta_{klp}(r,\phi,z)$ is the phase function.  We have for the amplitude function
 \beq
U_{klp}(\rho,\phi,z)={\cal U}_{k00} \frac{C_{lp}}{(1+z^2/z_R^2)^{1/2}}
\left(\frac{\sqrt{2}\rho}{{w(z)}}\right)^{|l|}L^{|l|}_p\left(\frac{2\rho^2}{ {w}^2(z)}\right) e^{-\rho^2/{w}^2(z)}. \label{envelope}
\eeq
Here $L^{|l|}_p$ is the associated Laguerre polynomial; ${\cal U}_{k00}$ is the amplitude for a corresponding plane wave of 
wavevector $k$;  $C_{lp}$ is a constant and $w(z)$ is the beam waist at position $z$ such that 
$w^2(z) = 2(z^2+z_R^2)/kz_R$, where $z_R$ is the Rayleigh range.  For the phase function we have
\beq
\Theta_{klp}(z)=kz+l\phi+\Theta_{Gouy}+\Theta_{curv}+i\omega t,\label{guoy}
\eeq
where 
\beq
\Theta_{Gouy}=-(2p+|l|+1) \tan^{-1}(z/z_{R});\;\;\;\;\;\;\;\Theta_{curv}=\frac{k\rho^{2}z}{2(z^{2}+z_{R}^{2})}.
\eeq   
The first term in the phase function is the usual term representing plane wave propagation with axial wavevector $k$ and the second term is the azimuthal phase which gives rise to $l$ intertwined helical wavefronts and is the basis for the OLM contents of the beam.  The third term is the Gouy phase and the final term enters as a phase contribution due to the variation of the beam curvature with both $\rho$ and $z$. 

\section{Optical forces on two-level atoms}
We focus on the interaction of a two-level atom of transition frequency $\omega_0$ and dipole moment $\mu$ with the LG light of frequency $\omega$ and electric field as describe above. In the steady steady such an atom is subject to position- and velocity-dependent forces.  A moving
atom in the field of a single LG beam experiences two forces: a scattering force and a dipole force
\beq 
{\bf F}_{klp}={\bf F}^{scatt}_{klp}+{\bf F}^{dipole}_{klp},
\eeq
where ${\bf F}^{scatt}_{klp}$ is the scattering force 
\beq
{\bf F}^{scatt}_{klp}({\bf r,v)}=\frac{\hbar\Gamma}{4}\Omega^{2}_{klp}({\bf
r})\left(\frac{{\bm {\nabla}}\Theta_{klp}({\bf
r})}{\Delta_{klp}^{2}({\bf r,v})+\Omega^{2}_{klp}({\bf
r})/2+\Gamma^{2}/4}\right),\label{48}
 \eeq 
and ${\bf F}^{dipole}_{klp}({\bf r,v})$ is the dipole force
 \beq
{\bf F}^{dipole}_{klp}({\bf r,v})=-\frac{1}{2}\hbar
\Omega_{klp}({\bf r}){\bm
{\nabla}}\Omega_{klp}\left(\frac{\Delta_{klp}({\bf
r,v})}{\Delta_{klp}^{2}({\bf r,v})+\Omega^{2}_{klp}({\bf
r})/2+\Gamma^{2}/4}\right),\label{49}
\eeq
where $\Omega_{klp}$ is the Rabi frequency and $\Delta_{klp}({\bf r,v})$ is the position- and
velocity-dependent detuning.  We have
\beq
\Omega_{klp}=\left|\frac{{\bn {\mu}}\cdot {{\bf E}}_{klp}}{\hbar}\right|;\;\;\;\;\;\Delta_{klp}({\bf r,v})=\Delta_{0}-{\bf v.}{\bm {\nabla}}\Theta_{klp}({\bf r,v}),
\eeq
where $\Delta_0=\omega-\omega_0$. The dipole force is derivable in terms of the gradient of the dipole potential as follows \cite{Babiker2018}
\beq
{\bf F}^{dipole}_{klp}({\bf r,v})=-{\bn {\nabla}}V^{dipole}_{klp}({\bf r,v}),
\eeq
where
\beq
V^{dipole}_{klp}({\bf r,v})=\frac{1}{2}\hbar
{\Delta}_{klp}({\bf r,v})\ln{\left[1+\frac{\Omega_{klp}^{2}({\bf
R})/2}{\Delta_{klp}^{2}({\bf r,v})+\Gamma_{klp}^{2}/4}\right]}. 
\eeq
Both the scattering force and dipole potential are well known when using ordinary (i.e. non-vortex) laser light in the context of atom cooling and trapping.  The
scattering force is a net frictional force responsible for optical molasses, and the dipole potential traps the atom in
regions of extremum light intensity.

\begin{figure}
\begin{center}\vspace{1cm}
\includegraphics[width=0.8\linewidth,angle=0.0]{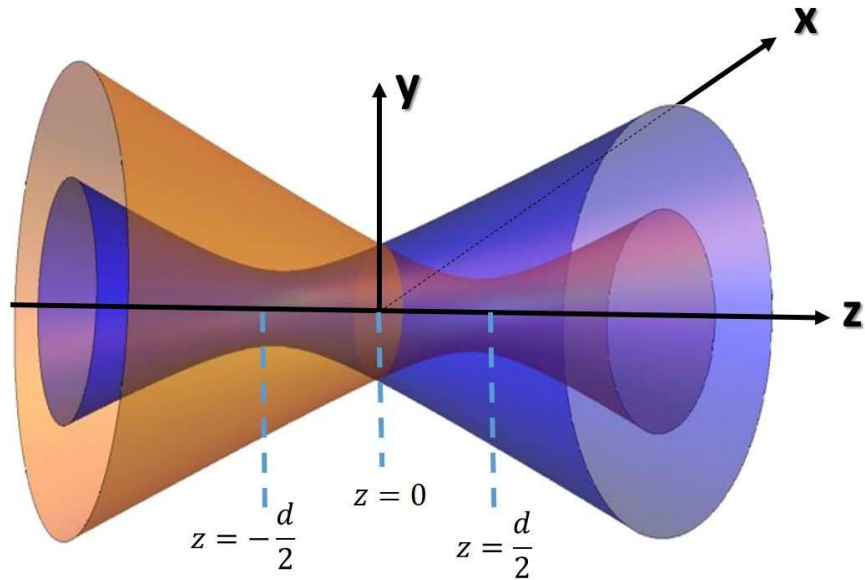}
\caption{Schematic representation of the shifted counter-propagating LG beams where beam 1 is taken to propagate along the positive z-direction with its focal plane positioned  at $z = -d/2$ while  beam 2 is propagating along the negative z-direction with its focal plane at $z= + d/2$.  We also assume that the two beams have the same polarisation ${\bn {\hat {\epsilon}}}$.}
\end{center}
\end{figure}
\vspace{1cm}
\section{Axially-shifted counter-propagating LG beams}
The physical system involves two counter-propagating LG beams, as shown schematically in Fig. 1 where beam 1 is taken to propagate along the positive z-direction with its focal plane positioned  at $z = -d/2$ while beam 2 is propagating along the negative z-direction with its focal plane at $z= + d/2$.  We also assume that the two beams have the same polarisation ${\bn {\hat {\epsilon}}}$.  We consider the case where both beams have low intensity with low winding number $l$ and at near resonance.  The influence of the weak light beams on a two-level atom in the region between the focal planes is the sum of individual forces arising from the two beams. The beams we are dealing with are doughnut beams for which $p=0$ and the same winding number $l$. Then the Gouy phase is negligible and we also drop the curvature phase term in this treatment. For ease of notation, we drop the mode label ${klp}$ and the superscripts defining the optical forces $scat$ and $dipole$.
At near-resonance, where $Delta_0$ is small, the dipole force of a two-level atom is small and the scattering force dominates. 

We seek to identify the first feature introduced by the shifting of the beam focal planes, so we shall consider velocity-independent i.e. static optical forces.  Setting ${\bf v}=0$, there are two scattering forces on the atom arising from the two beams which we can conveniently describe using cylindrical polar coordinates ${\bf r}=({\bn {\rho}},z)$, with ${\bn {\rho}}=(\rho,\phi)$ the in-plane position vector component in polar coordinates. We have
\beq
{\bf F}_1({\bf r})=\frac{\hbar\Gamma}{4}\frac{\Omega^2({\bn {\rho}},z+d/2)}{\Delta_0^2+\Gamma^2/4+\Omega^2({\bn {\rho}},z+d/2)/2}\left[k{\bn {\hat z}}+\frac{l}{\rho}{\bn {\hat \phi}}\right],
\eeq                                     
\beq
{\bf F}_2({\bf r})=\frac{\hbar\Gamma}{4}\frac{\Omega^2({\bn {\rho}},z-d/2)}{\Delta_0^2+\Gamma^2/4+\Omega^2({\bn {\rho}},z-d/2)/2}\left[-k{\bn {\hat z}}+\frac{l}{\rho}{\bn {\hat \phi}}\right],
\eeq 
Recall that the mode labels have been suppressed for ease of notation and $\Omega$ is the Rabi frequency and $\Gamma$ is the inverse lifetime of the excited state of the two-level atom.

The total force acting on the atom is the sum 
\beq
{\bf F}({\bf r})={\bf F}_1({\bf r})+{\bf F}_2({\bf r}).
\eeq

We are interested in the region between the focal planes as a trapping region for atoms and by symmetry we expect the total force to be the same with reference to the centre of the system, located at $z=0$.
Hence we seek to explore the small $|z|$ region using Taylor expansion.  We have the leading order
\beq
{\bf F}_1({\bf r})\approx \frac{\hbar\Gamma}{4}\left[k{\bn {\hat z}}+\frac{l}{\rho}{\bn {\hat \phi}}\right]\left\{Q_+(z=0)+z\left(\frac{\partial Q_+}{\partial z}\right)_{z=0}\right\}.
\eeq
where 
\beq
Q_+=\frac{\Omega^2({\bn {\rho}},z+d/2)}{\Delta_0^2+\Gamma^2/4+\Omega^2({\bn {\rho}},z+d/2)/2}
\label{kew+}
\eeq

A similar Taylor expansion can be carried out for ${\bf F}_2$ and we note that at $z=0$ we have $Q_+=Q_-$ and $\Omega^2_{d/2}=\Omega^2_{-d/2}$.  Adding the contributions from ${\bf F}_1$ and ${\bf F}_2$ to leading order, we find
\beq
{\bf F}({\bf r})=\frac{\hbar\Gamma k}{2}\left(\frac{\partial Q_+}{\partial z}\right)_{z=0}z{\bn {\hat z}}+\frac{2l}{\rho}Q_+(z=0){\bn {\hat \phi}}.
\eeq
where 
\beq
\fl
\left(\frac{\partial Q_+}{\partial z}\right)_z=-d\frac{(\Delta_0^2+\Gamma^2/4)\Omega^2(z=d/2)}{[\Delta_0^2+\Gamma^2/4+\Omega^2(z=d/2]^2}\left\{\frac{(|l|+1)(z_R^2+d^2/4)-2\rho^2z_R^2/w_0^2}{(z_R^2+d^2/4)^2}\right\}.
\eeq
We see that the axial total force is a quasi-restoring force centred  at $z=0$ and can be written as 
\beq
{\bf F}_z=-{\cal K}z,
\eeq
where ${\cal K}$ is the spring `constant', which here is weakly dependent on $\rho$. We have 
\beq
\fl
{\cal K}=\frac{\hbar\Gamma k}{2}d\frac{(\Delta_0^2+\Gamma^2/4)\Omega^2(z=d/2)}{[\Delta_0^2+\Gamma^2/4+\Omega^2(z=d/2]^2}\left\{\frac{(|l|+1)(z_R^2+d^2/4)-2\rho^2z_R^2/w_0^2}{(z_R^2+d^2/4)^2}\right\}.\label{young}
\eeq 
The total axial force exhibits a quasi-harmonic trapping potential between the foci which is centred at $z=0$, where the two beams supply the same intensity with a combined single doughnut ring. This central doughnut ring is of radius $\rho_0$ and it is at an axial distance $d/2$ from the focal planes.  We have
$\rho_0=\sqrt{w^2(z=d/2)l/2}$. Substituting for $w(z=d/2)$ and noting that $w^2_0=2z_R/k$ we find $\rho_0=w_0\sqrt{l/2}\sqrt{1+\frac{d^2}{4z_R^2}}\label{rho0}$. 
Note that the spring 'constant' ${\cal K}$ depends on the radial coordinate $\rho$, as given in Eq.(\ref{young}).  This takes a simpler form at the radial position $\rho=\rho_0$, i.e. at the central doughnut ring. Substituting for  $\rho=\rho_0$  we  find that the expression between the curly brackets in Eq.(\ref{young}) becomes equal to unity and we find

\begin{equation}
{\cal K}_0={\cal K}(\rho_0,z=0)=\frac{\hbar\Gamma k}{2}d\frac{(\Delta_0^2+\Gamma^2/4)\Omega^2(z=d/2)}{\left\{[\Delta_0^2+\Gamma^2/4+\Omega^2(z=d/2]^2(z_R^2+d^2/4)\right\}}.
\end{equation}

This is a constant which depends on the system parameters, most notably, the separation $d$.  Figure 2 displays the variations of the spring constant ${\cal K}_0$ with $d$ for a typical set of parameters, as stated in the caption to the figure.
\begin{figure}
\begin{center}\vspace{1cm}
\includegraphics[width=0.6\linewidth,angle=-90.0]{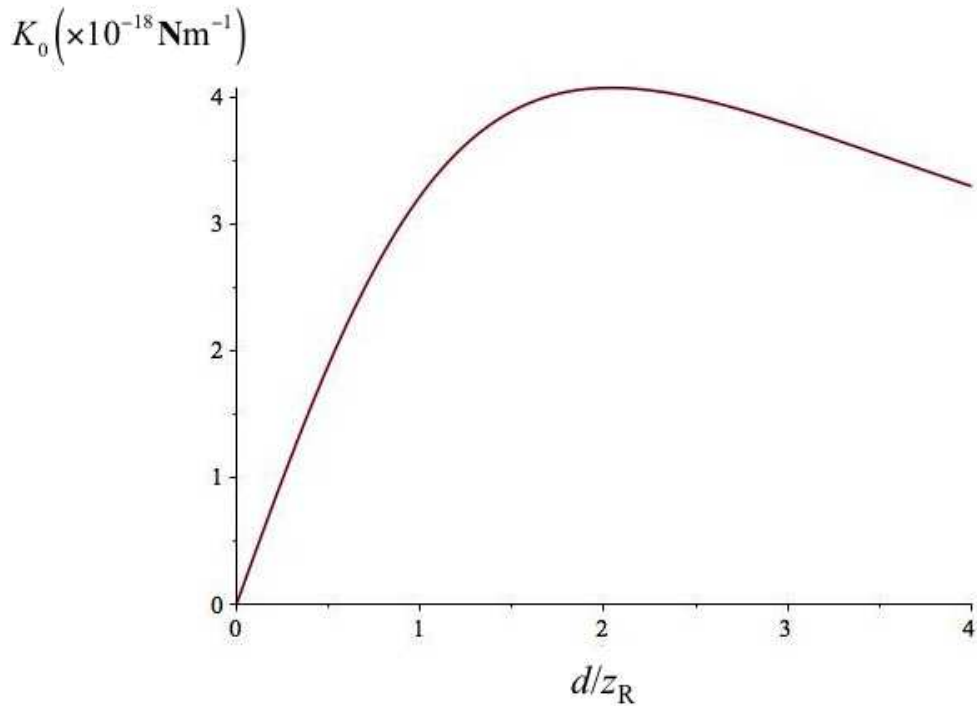}
\caption{The variation of the spring constant ${\cal K}_0$ with $d$ for sodium atoms with the transition $^2S_{1/2} - ^2P_{3/2}$ of wavelength $\lambda_0=589.16\;$ nm and $\Gamma=2\pi\times 10.01\;$MHz.  Other parameters are as follows: $\Delta_0=0.5\Gamma$, $w_0=8\mu$m$, \Omega_0=\Gamma$, where $\Omega_0$ is the Rabi frequency corresponding to a field with an amplitude ${\cal U}_{k00}$ appearing in Eq.(\ref{envelope}).}
\end{center}
\end{figure}
\vspace{1cm}

Since ${\cal K}_0$ is a constant, we may now talk about a true harmonic potential $V_0(z)$ associated with the axial scattering force and so write
\beq
V_0(z)=\frac{1}{2}{\cal K}_0z^2,
\eeq 
so that $F_z(z)=-{\bn {\nabla}}V_0(z)$. The physical situation is now clear in that the atoms will be trapped within the axial potential well with a minimum at the plane $z=0$.  The azimuthal component at $\rho_0 $, given by ${\bf F}_{\phi}=(2l/\rho_0) Q_+(z=0){\bn {\hat \phi}}$ of the scattering force acts to rotate the atoms by the light induced torque given by 
\beq
{\bf T}={\bf r}\times{\bf F}_{\phi}=2lQ_+(\rho_0,z=0){\bn {\hat z}}.
\eeq
where $Q_+$ is given by Eq.(\ref{kew+}).  

The above treatment is applicable in the low intensity limit and also does not take into account the situations involving focused beams with significant contributions from the Gouy phase and the curvature phase.  Our next task is to explore the general case where interference is an important ingredient along with moderate focussing, so that we are still within the paraxial regime and the full LG formalism described at the outset is applicable.

\section{Mixing phases and amplitude functions}
We now consider the general case where the beams are sufficiently intense and the winding number $l$ is sufficiently large for the curvature effects, including the Gouy phase to come into play. With the two beams having the same polarisation ${\bn {\hat {\epsilon}}}$,  the total field is simply the sum of the field vectors.

Once again, for ease of notation we suppress the LG mode labels $klp$ and restore these when the need arises.  For beams 1 and 2 we write, 
\beq
{\bf E}_{1}=U_{1}e^{i\Theta_{1}}{\bn {\hat {\epsilon}}}; \;\;\;\;\;\;{\bf E}_{2}=U_{2}e^{i\Theta_{2}}{\bn {\hat {\epsilon}}},
\eeq
where the amplitude functions $U_1$ and $U_2$ and phase function $\Theta_1$ and $\Theta_2$ are appropriate for LG  beams of the forms in Eqs.(\ref{envelope}) and (\ref{guoy}).  

Since the beam polarisations are the same we write for the total field
\beq
{\bf E}=Ue^{i\Theta}{\bn {\hat {\epsilon}}},
\label{totalE}
\eeq
where $U$ and $\Theta$ are the total amplitude function and the total phase function of the interfering beams.  The evaluation of these functions proceeds as follows
\beq
Ue^{i\Theta}=\left(U_1e^{i\Theta_1}+U_2e^{i\Theta_2}\right).
\eeq
Writing the complex exponentials on the right-hand side in terms of sine and cosine functions of $\Theta_1$
and $\Theta_2$, followed by separation of real and imaginary part we are then able to write straight forwardly for the total amplitude function
\beq
U=\left\{U_1^2+U_2^2+2U_1U_2\cos(\Theta_1-\Theta_2)\right\}^{1/2}.
\label{totalamplitude}
\eeq
For the total phase function we have
\beq
\Theta=\tan^{-1}\left\{\frac{U_1\sin\Theta_1+U_2\sin\Theta_2}{U_1\cos\Theta_1+U_2\cos\Theta_2}\right\}.
\label{totalphase}
\eeq
Equations (\ref{totalamplitude}) and (\ref{totalphase}) succinctly represent the interference of any two LG beams.
The total amplitude function shows interference effects residing in the cosine function involving the phase difference $\Theta_1-\Theta_2$.  On the other hand the total phase involves the amplitude functions.  As we show below this inter-mixing of the amplitudes and phases in the total field turns out to be the source of interesting effects in the context of LG beams.

\section{Axially-shifted counter-propagating doughnut beams}
Our main concern here is to explore the specific case of shifted counter-propagating LG modes and proceed to determine their intensity and phase distributions and their  influence on the atoms with which they interact at near resonance. For simplicity we continue to focus on doughnut beams $LG_{l_1,0}$ and $LG_{l_2,0}$.

As before, we assume that the two beams have the same frequency $\omega$ and we take the focal plane of beam 1 to be situated at the point $z=-d/2$ and that of beam 2 to be situated at $z=d/2$, with the origin of coordinates situated at z=0, as shown in Fig.1. we have for the amplitude and phase of beam 1
\beq 
U_1=U(z\rightarrow z-d/2);\;\;\;\;\Theta_1=\Theta(z\rightarrow z-d/2).
\eeq
where $U$ and $\Theta$ are adaptable from by Eqs.(\ref{envelope}) and (\ref{guoy}). Similarly, the expressions appropriate for beam 2 can be written where $z\rightarrow z+d/2$. 

\subsection{Total phase function}

In writing the phase function of beam 2 we must take into account that this beam is travelling along -z in addition to the shift of focal plane.  
Once the expressions for $U_1, U_2, \Theta_1$ and $ \Theta_2$ have been determined the the total phase function follows by direct substitution in Eq.(\ref{totalphase})

\subsection{Total amplitude function and power density}

The total amplitude function is given by Eq.(\ref{totalamplitude}) where we need to substitute for $U_1$ and $U_2$, but we also need to evaluate the phase difference. We find since the beams have the same frequency $\omega$
\beq
\Theta_1-\Theta_2=2kz+(l_1+l_2)\phi+\Delta\Theta_{Gouy}+\Delta\Theta_{curv},\label{thetadiff}
\eeq
where 
\beq
\Delta\Theta_{Gouy}=-(|l_1|+1)\tan^{-1}\left(\frac{z-d/2}{z_R}\right)-(|l_2|+1)\tan^{-1}\left(\frac{z+d/2}{z_R}\right)\label{deltagouy}.
\eeq
For counter-propagating doughnut beams with $l_1=l_2=l$ this simplifies to
\beq
\Delta\Theta_{Gouy}=-(|l|+1)\tan^{-1}\left\{\frac{2zz_R}{z_R^2-(z^2+d^2/4)}\right\}
\eeq
For $\Delta\Theta_{curv}$ we find 
\beq
\Delta\Theta_{curv}=\frac{k\rho^2d}{2(z^2+z_R^2)}.
\eeq

\begin{figure}
\begin{center}\vspace{1cm}
\includegraphics[width=0.8\linewidth,angle=0.0]{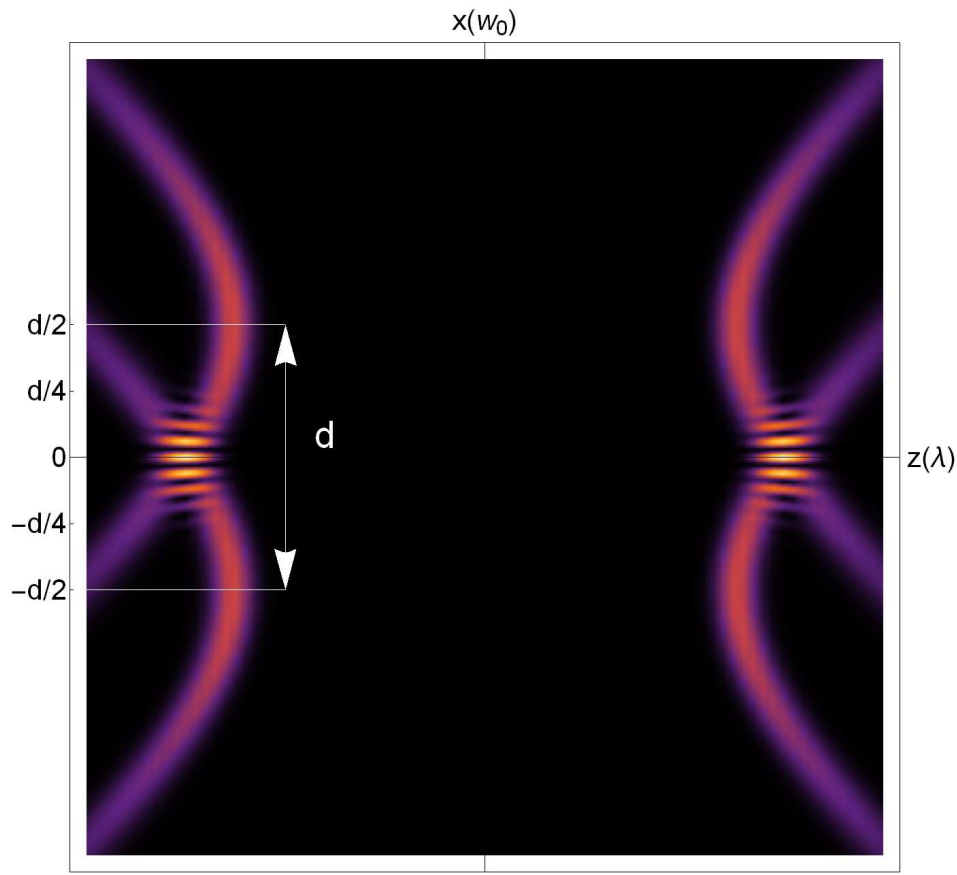}
\caption{The intensity profile of the interference pattern of two counter-propagating doughnut LG beams.  The pattern is shown projected on the $x$-$z$ plane at $y=0$ with the focal planes at $z=-d/2$ and $z=d/2$, with the centre of this cylindrically symmetric system located at $z=0$.  The two sides of the interference fringes form the cross-section of the concentric rings, shown in figure \ref{Fig4}.  The parameters used here are as follows: $l_1=l_2=80$, $w_0=6\lambda$ and $d=24w_0$. These paraxial parameters are consistent with those adopted by Chu {\it et al.}\cite{Chu1986}}
\label{Fig3}
\end{center}
\end{figure}

\begin{figure}
\begin{center}\vspace{1cm}
\includegraphics[width=0.8\linewidth,angle=0.0]{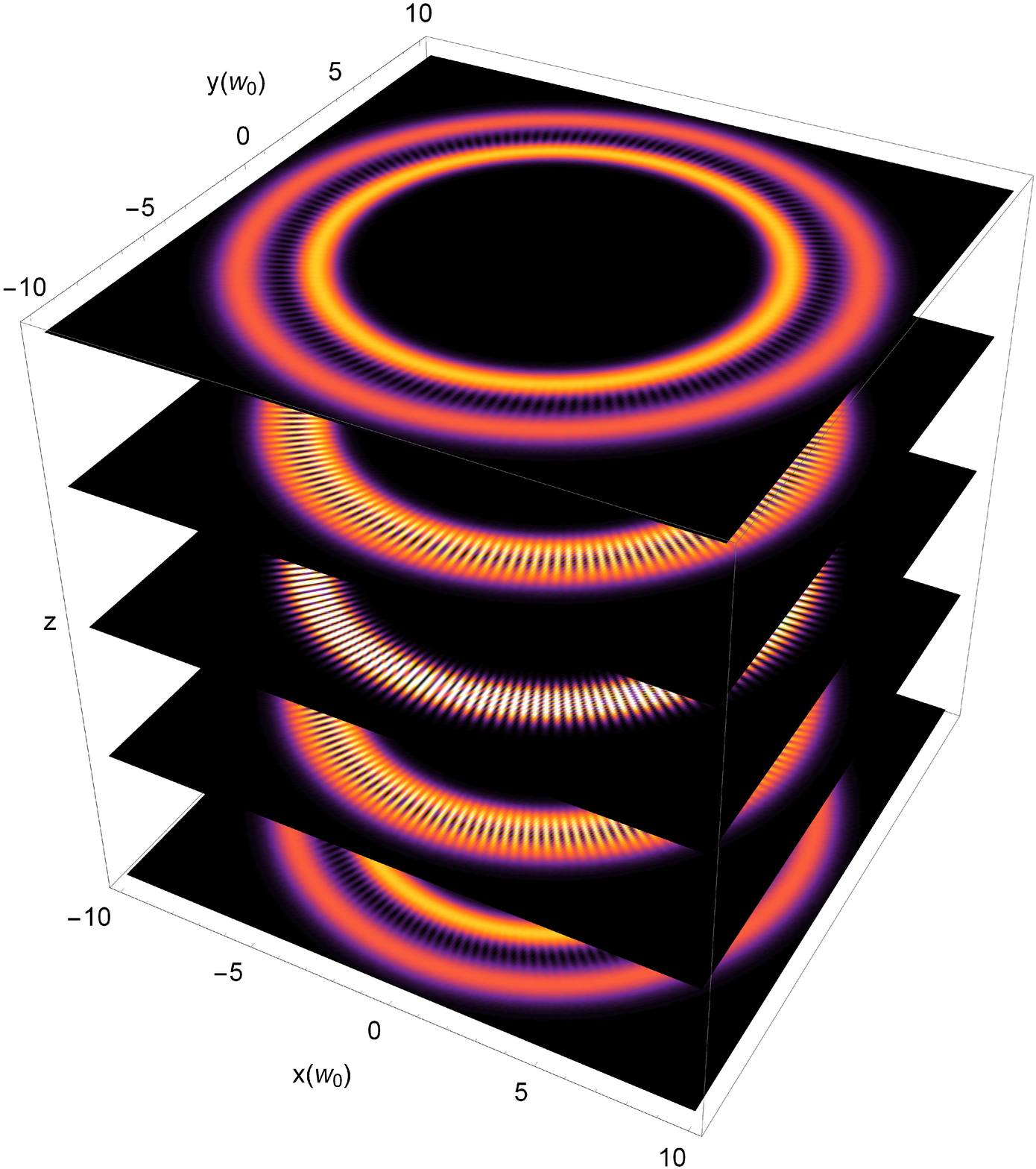}
\caption{The interference intensity pattern projected on the $xy$ planes in the region between the focal planes.  The parameters are as in Fig. \ref{Fig3}. This figure shows a central rings at $z=0$ and double rings on either sides of this central ring.  The modulation is due to the cosine of the phase difference $\cos{(\Theta_1-\Theta_2)}$ entering the total amplitude function equation (\ref{totalamplitude}).}
\label{Fig4}
\end{center}
\end{figure}

The explicit form of the phase difference $\Theta_1-\Theta_2$ is
\beq
\Theta_1-\Theta_2=2kz+2l\phi-(|l|+1)\tan^{-1}\left\{\frac{2zz_R}{z_R^2-(z^2+d^2/4)}\right\}+\frac{k\rho^2d}{2(z^2+z_R^2)}.
\eeq
This enables the total power density distribution to be evaluated involving realistic parameters as described in the caption to Fig. \ref{Fig3} and \ref{Fig4}.

The superposition is seen to give rise to a standing wave in the form of a finite ring lattice of coaxial intensity rings spanning the axial region between the focal planes.  The modulation in the power density pattern is governed by the cosine of the phase difference $\cos(\Theta_1-\Theta_2)$ entering the total amplitude function.

\subsection{Effects of frequency shift}
In addition to the beams being shifted in space along the common axis, it is possible for their frequencies to differ slightly by $\Delta\omega$ in which case the total amplitude function becomes
\beq
U=\left\{U_1^2+U_2^2+2U_1U_2\cos(\Theta_1-\Theta_2-\Delta k z-\Delta\omega t\right\}^{1/2},
\label{totalamplitude2}
\eeq
where $\Delta\omega=\omega_1-\omega_2$ and $\Delta k=k_1-k_2$.  The $\Delta k$ term is responsible for axial beating/dephasing effects exhibited by the envelope function in the z-direction.  These are typically negligible  for laser beams as the dephasing length $2\pi/\Delta k$ is typically much longer than the longitudinal coherence length of the laser beam. Since the argument of the cosine function in Eq.(\ref{totalamplitude2})is now time-dependent leading to optical Ferris wheels and lift.

\subsection{Optical Ferris wheels and lifts}
The time dependence arising due to the frequency shift $\Delta\omega$ means that the interference pattern now moves with time.  A given plane, for instance the symmetry plane $z=0$, can be shown to shift with time.  There is azimuthal motion in every doughnut ring in the interference pattern between the focal planes which moves at an angular frequency $v_{\phi}$ given by
\beq
v_{\phi}=\frac{d\phi}{dt}=\frac{\Delta\omega}{2l}.
\eeq
Thus, in the context of shifted beams this is a manifestation of the so-called optical Ferris wheel described by Franke-Arnold {\it et al.} \cite{Franke-Arnold2007}.  An additional effect arising from the introduction of a frequency difference $\Delta\omega$ is a translation of the interference pattern along the $z$-axis at a speed $v_z$ given by
\beq
v_z=\frac{dz}{dt}\approx \frac{\Delta\omega}{2k},
\eeq
where the approximation is necessitated by having dropped the Gouy phase and curvature phase terms in the argument of the cosine function.  This is a reasonable approximation for the case of beams with large Rayleigh range $z_R$, and/or small winding number $l$.  In the present context the motion of the interference pattern along the $z$-axis is a manifestation of the so-called lifting effect, or the conveyor belt effect. 

\subsection{Radial shifts in double rings}

As Fig. \ref{Fig4} shows, the interference intensity pattern consists of a set of double doughnut rings, with a single central ring at $z=0$.  On either side of this brightest ring there are rings separated by a radial distance $\delta\rho$, depending on the position $z$.  We see that there is a double ring at $z=(d/2-\delta)$ where $\delta$ is the fringe separation. One of the double rings is a distance equal to $d/2-\delta$, from the focal plane of beam 1 while the second ring is a distance of $d/2+\delta$ from the focal plane of beam 2.  Thus the radii of the rings are given by 
\beq
\fl
w_1=w_0\sqrt{|l|/2}\sqrt{[1+(d/2-\delta)^2/z_R^2]};\;\;\;\;\;w_2=w_0\sqrt{|l|/2}\sqrt{[1+(d/2+\delta)^2/z_R^2]}.
\eeq
The radial separation is the difference.  We have
\beq
\Delta\rho=w_0\sqrt{|l|/2}\left(\sqrt{1+(d/2+\delta)^2/z_R^2}-\sqrt{1+(d/2-\delta)^2/z_R^2}\right).
\eeq
For small $d$ or large $z_R$ this radial separation is approximately $\Delta\rho\approx w_0\sqrt{2|l|}(d\delta/z_R^2)$.  This provides the possibility of choosing the parameters in such a manner that the two component rings can be made to be very close to each other.  If  atoms are trapped in these rings, they can tunnel between the rings for sufficiently small radial separations. For $l$ small this could be advantageous when compared with schemes where LG beams with the same $l$, but $p=1$ are examined for atom tunnelling between ring traps.  In that case, the inner ring is separated from the outer one by a radial distance of $w_0$, while in the current case the factor $\alpha=\sqrt{2|l|}(d\delta/z_R^2)$ may be arranged to be less than unity.

\section{Conclusions}

The system we have considered here, namely that involving counter-propagating LG beams with axially shifted focal planes, as far as we know,  has not been explored before in the context of twisted light.  This requires the introduction of a spatial separation $d$ between the two focal planes of the counter-propagating LG beams, giving  rise to novel intensity and phase distributions. We have concentrated on the case in which the LG beams are doughnut beams and both beams are assumed to be linearly polarized in the same direction so that when the total field is constructed as the sum of the field vectors, a total amplitude function and a total phase function can be defined and are found to be intertwined, due to spatial variations in the individual beams.  The properties of the total fields can be controlled by changing the separation d and the curvature of the beams which alters the useful length of the interference pattern. In addition to the intensity distribution, which is shown to consist of a finite ring lattice, significant other properties are identified due to variations in the total phase function and its gradient.  We have also shown that when the frequencies of the beams differ slightly the interference pattern becomes a set of rotating Ferris wheels which also move axially between the focal planes, identified as a conveyor belt in this context.  

The summary of our main results is as follows: (1) We have identified and quantified a new trapping mechanism of two-level atoms that arises due to the scattering force, which vanishes in the more common cases in which $d=0$.  We feel that this result is significant, not only because of the novelty of the finding, but also because it entails an all-optical trapping mechanism that would be an alternative to other more involved trapping mechanisms such as magneto-optical trapping. (2) The one-dimensional case considered here highlights the role of the Gouy phase and curvature in that they control the spacing and number of the rings in the interference pattern.  We have emphasised the role of the phase difference $\Theta_1-\Theta_2$ in controlling various features including the number and spacing (both radially and axially) of the rings. (3) We have pointed out and quantified effects that would be of interest to experimentalists, including atom trapping by both scattering and dipole forces and have examined regions for atom tunnelling between double rings.

\section*{Acknowledgements} KK is grateful to TUBITAK, Turkey, for financial support during a sabbatical leave at the University of York, UK, where this work was initiated. This project was supported by King Saud University, Deanship of Scientific Research, College of Science Research Centre.

\section*{ORCID IDs}

K Koksal https://orcid.org/0000-0001-8331-9380\\
Vasileios E Lembessis  https://orcid.org/0000-0002-2000-7782\\
J Yuan  https://orcid.org/0000-0001-5833-4570\\
M Babiker  https://orcid.org/0000-0003-0659-5247

\end{document}